\documentclass[%
 reprint,
superscriptaddress,
 amsmath,amssymb,
 aps,
]{revtex4-1}

\usepackage{graphicx}
\usepackage{dcolumn}
\usepackage{bm}
\usepackage{hyperref}
\usepackage{fancyref}
\usepackage{braket}
\usepackage{natbib}

\usepackage[printwatermark]{xwatermark}
\usepackage{xcolor}

\begin{document}

\preprint{APS/123-QED}

\title{Condensate Dynamics with Non-Local Interactions
}

\author{Erik W Lentz}
\affiliation{Institut f\"ur Astrophysik, Georg-August Universitat G\"ottingen, 
                 Goettingen, Deutschland 37707;       
		  {\tt erik.lentz@uni-goettingen.de}}
\affiliation{Physics Department, University of Washington,
                 Seattle, WA 98195-1580;       
		 {\tt lentze@phys.washington.edu, ljrosenberg@phys.washington.edu}}
\author{Thomas R Quinn}
\affiliation{Astronomy Department, University of Washington, 
                 Seattle, WA 98195-1580;       
		  {\tt trq@astro.washington.edu}}
\author{Leslie J Rosenberg}
\affiliation{Physics Department, University of Washington,
                 Seattle, WA 98195-1580;       
		 {\tt lentze@phys.washington.edu, ljrosenberg@phys.washington.edu}}

\date{\today}

\begin{abstract}
Systems of identical particles possessing non-local interactions are capable of exhibiting extra-classical properties beyond the characteristic quantum length scales. This letter derives the dynamics of such systems in the non-relativistic and degenerate limit, showing the effect of exchange symmetry and correlations on structure both in and out of equilibrium. Such descriptions may be crucial to understanding systems ranging from nuclei to dark matter. Appropriate limits for restoring the mean-field description are also discussed.
\end{abstract}

\pacs{Valid PACS appear here}

\maketitle


\section{\label{sec:Introduction}Introduction}

Identical particles with non-local interactions are common in non-relativistic quantum mechanics. Many intriguing systems in atomic and nuclear physics, condensed matter, astrophysics, and cosmology may fundamentally be understood by many-body quantum mechanics subject to non-local and sometimes infinite-range inter-particle forces \cite{CMMBbook,Dobaczewski2011,Fiolhais2003,Marsh2016}. 

Degenerate systems are particularly interesting as their coherence can bring quantum behavior into prominence. The effects of degeneracy often permits significant simplification and, on occasion, almost innumerable possible configurations are able to be described straightforwardly with only a single degree of freedom.

Mean field theory (MFT) is among the most popular approaches to studying such systems. Often relying on conditions of separability among bodies, diffusion, or other means of ensuring uncorrelated motion, such models can describe entire systems with very simple equations, such as Vlassov's \cite{Vlasov1968,Eby2016,Levkov2017}. However, such conditions restrict solutions to a small portion of the system's domain, with no guarantee that the sub-space intersects the region of interest. These constraints may then influence the dynamics of the system in an un-physical way.

\begin{figure}[h]
\begin{center}
\includegraphics[width=9cm]{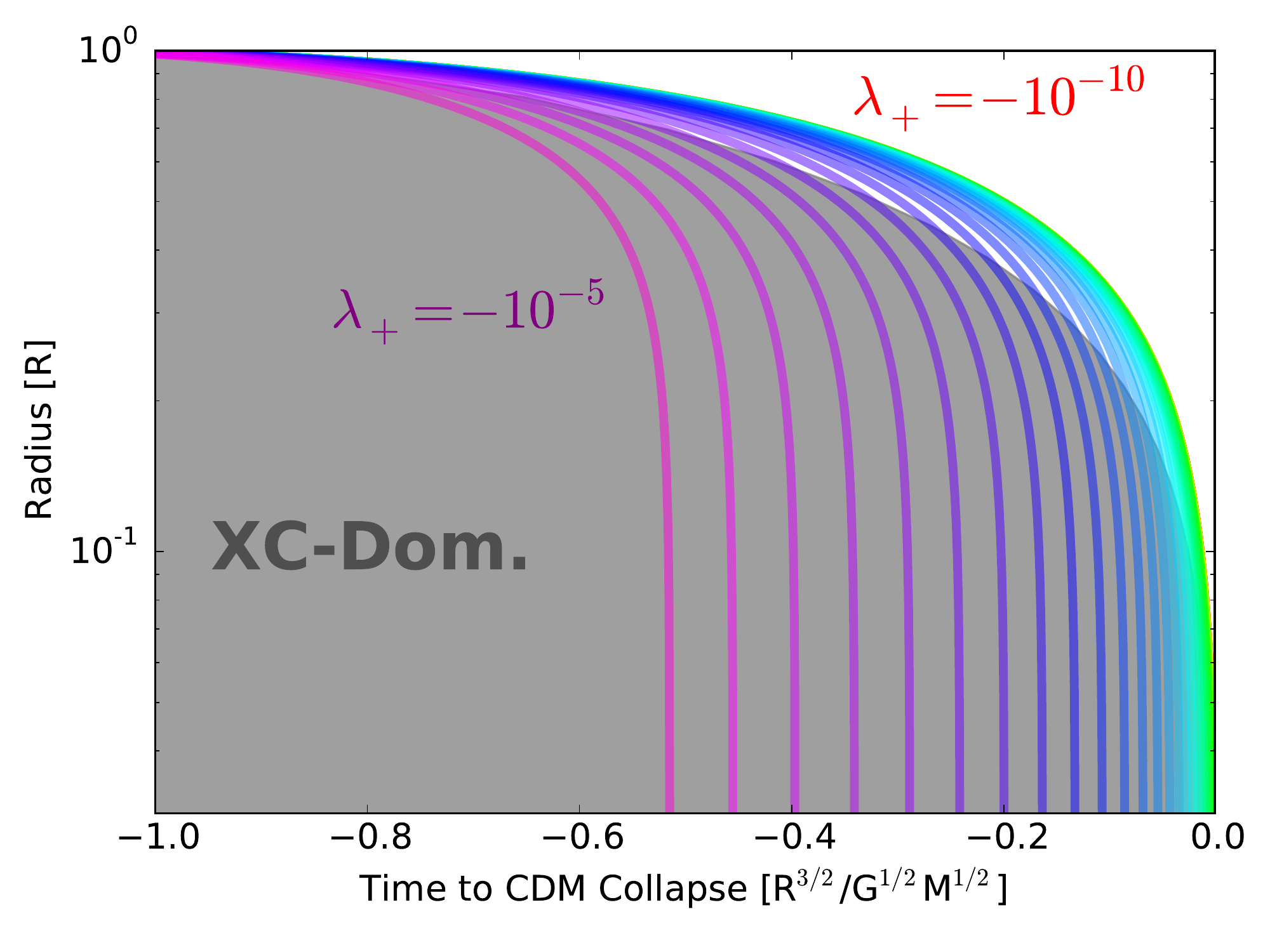}
\caption{\textbf{Demonstration of XC dynamics on gravitational collapse}:  Radial trajectories of collapsing condensed self-gravitating spherical shells of QCD axion dark matter. Shells start from rest. The color gradient is logarithmic in XC strength $\lambda_+$, which is left as a free constant parameter, running from the near-classical limit of $\lambda_+=-10^{-10}$ at the top to the much stronger contribution of $\lambda_+=-10^{-5}$ at the bottom, evaluated in units of R$^{3/2}$M$^{1/2}$G$^{-1/2}$. The XC physics obviously alters the infall of the shell, generally increasing the attraction and making the collapse more violent. The locations where XC physics dominates over classical gravity are shaded in gray. Details of the model are covered in Section \ref{sec:Example}.}
\label{sph_shell_coll}
\end{center}
\end{figure}

Identical-particle quantum mechanics has its own constraints, principally permutation symmetry or anti-symmetry according to the nature of the intrinsic spin degrees of each particle. The permutation constraint on these boson (symmetric) and fermion (anti-symmetric) systems are also expected to influence dynamics \cite{Beinke2018}. The scale independence of this constraint suggests the range of this dynamics change may extend beyond the standard quantum length scales. One may expect such effects to be most visible in the system's degenerate phases. Together, the degeneracy and permutation influences are referred to as exchange-correlation (XC) effects.

This letter shows that the dynamics of non-locally interacting identical particles, both bosonic and fermionic, have a concise description beyond the de Broglie scale in the degenerate limit. The description presents exchange-correlation effects naturally via an extremal inter-particle correlation function. The standard mean-field description is recovered in the homogeneous limit.  An example calculation is performed on a self-gravitating degenerate Bose fluid.

\section{\label{sec:Derivation}Derivation}

The models considered here evolve under a many-body Schr\"odinger equation, dictated by the Hamiltonian operator $\hat{H}$, which for simplicity contains no spin dependence. Further, the solution space is to be restricted to states that can be written as a product of spatial and non-spatial degrees, where the non-spatial degrees are in symmetric form. The Schroedinger equation expressed in the spatial basis then takes the form
\begin{equation}
i \hbar \partial_t  \Psi \left( \vec{x}_1,...,\vec{x}_N;t \right) = \hat{H} \Psi \left( \vec{x}_1,...,\vec{x}_N;t \right)
\end{equation}
with Hamiltonian
\begin{equation}
\hat{H} = - \sum_i^N \frac{\hbar^2 \nabla^2_i}{2 m} + \sum_{i < j}^N \phi \left(\vec{x}_i-\vec{x}_j \right)
\end{equation}
where the first term is the canonical kinetic-energy contribution from each particle, and the second term represents the non-local inter-particle interactions, parameterized by the single, smooth, real potential function $\phi$. Only non-local central interactions are considered here. The exchange condition can now be written as 
\begin{align}
\Psi \left( \vec{x}_1,...,\vec{x}_i,...,\vec{x}_j,...,\vec{x}_N;t \right) &= \mp \Psi \left( \vec{x}_1,...,\vec{x}_j,...,\vec{x}_i,...,\vec{x}_N;t \right) \nonumber \\ 
& \forall i,j
\end{align}
for fermions and bosons respectively.

The domain of the Hamiltonian operator is contained in the Hilbert space $L^2(\Re^{3N})$, the time operator is spanned by the set of smooth complex functions, and together $L^2(\Re^{3N})\otimes C_0^1(\Re)$ spans the domain of the Schr\"odinger differential operator. The solution space is therefore isomorphic to
\begin{equation}
L^2(\Re^{3}) \otimes ... \otimes L^2(\Re^{3}) \otimes C_0^1(\Re)/S_N
\end{equation}
where $S_N$ is the (anti-)equivalence among the totally-symmetric group of particle permutations.

The center-of-mass solutions of this system are spanned by fully (anti-)symmetric combinations inter-particle correlators
\begin{align}
&\Psi \left( x,t \right)  \in \nonumber \\
&\left\{
	\begin{array}{ll}
		sp \left( \left\{ det \left( \prod_{ij } \psi_{\alpha_{ij}} \left(\vec{x}_i - \vec{x}_j,t \right) \right) \right\}_{\alpha \in A} \right)  & \mbox{(Fermion) } \\
		sp \left( \left\{ perm \left( \prod_{ij } \psi_{\alpha_{ij}} \left(\vec{x}_i - \vec{x}_j,t \right) \right) \right\}_{\alpha \in A} \right) & \mbox{(Boson) }
	\end{array}
\right. \label{cohstates}
\end{align}
where $sp()$ gives the linear span of a collection of functions, $perm()$ is the permanent operation, $det()$ is the determinant operation, $A$ is the set of many-body solution indexes, $\hat{\psi}_{a}$ is a solution to the single-body equation
\begin{equation}
i \hbar \partial_t \psi_{a} = - \frac{\hbar^2}{2 \mu} \nabla^2_{\left( \vec{x}_i - \vec{x}_j \right)} \psi_{\alpha} + \phi\left( \vec{x}_i - \vec{x}_j \right) \psi_{a}
\end{equation}
  and $\mu$ is a particle reduced mass and scales as $1/N$, thereby giving a naturally-extended potential length scale to inter-particle correlations. The inter-particle potential's length scale, the number of particles $N$, and the global reach of permutation (anti-)symmetry are therefore key to the development of the system's structure. This correlator form explores the solution space more efficiently than the single-particle expansion natural to Fock space representations, especially in the presence of highly-correlated interactions. The utility of this form lies in the implicit embedding of exact exchange symmetry and the effects of that constraint on possible actions of the system; this is crucial in the expression of physics beyond the mean-field. Only condensed configurations are considered for the remainder of this paper, where condensed here is taken to mean occupation of a single basis element of Exp. \ref{cohstates} state with $\alpha_{ij} = \alpha$ for bosons, and $N^2$ unique $\alpha_{ij}$ that are extremally correlated for fermions.

The dimensionality of the condensed system may be reduced by a combination
of the Wigner transform \citep{casas1990} and integration. The many-body Wigner
transform produces a distribution equation of motion that matches the
Liouville form in the super-de Broglie, small $\hbar$, limit
\begin{equation}
\partial_t f^{(N)} + \sum_i^N \frac{\vec{p}_i}{m} \vec{\nabla}_i f^{(N)} -\sum_{i < j}^N \vec{\nabla}_i \phi_{ij} \cdot \vec{\nabla}_{p_i} f^{(N)} = O(\hbar) \label{DFEOM}
\end{equation}
where $f^{(N)}$ is the N-body distribution function (DF) and $\phi_{ij} = \phi(\vec{x}_i-\vec{x}_j)$. 

Applying the Runge-Gross theorem \cite{Runge1984,TDDFT1984}, which proves the existence of an injective mapping from the potentials of a of many-body quantum mechanical system to single-body density given an initial many-body wave-function, reveals that the only important degree of freedom for the identical system is its density, $\rho$, which makes it possible to reduce dimensionality without loss of generality. This treatment aims for such a concise description by integrating out phase-spaces $2,...,N$ of Eqn. \ref{DFEOM}. This is straightforward for the first two terms, but the two-body interaction requires more consideration. The normal ``molecular chaos'' approach to distribution theory would lead one to take a near-uncorrelated form of the distribution $f^{(2)} = f^{(1)} \otimes f^{(1)} + g$. The correlation function of this form is often restricted to  $1/N$ scaling by local considerations of two-particle scattering \cite{Fokker1914,Kolmogoroff1931}. However, the condensed state implies no such scaling due to the global reach of the permutation condition. Instead, the extremal nature of the condensate is used to construct a correlation optimization problem.

We write $f^{(2)}$ in a form that makes the correlations between single-body DFs explicit
\begin{equation}
f^{(2)}(w_1,w_2,t) = \tilde{g} \cdot f^{(1)}(w_1,t) f^{(1)}(w_2,t) 
\end{equation}
where $w_i$ is the six-dimensional phase space of the i-th particle, and $\tilde{g}$ is a correlation function more reminiscent of the two-body correlators of the underlying wave equation. The correlation function $\tilde{g}$ is symmetric over both phase spaces, and one representation of $\tilde{g}$ that depends only on $f^{(1)}$ is
\begin{equation}
f_+ \equiv \frac{1}{2}\left(f^{(1)}(w_1,t) + f^{(1)}(w_2,t) \right)
\end{equation}
Condensed correlation functions can be shown to satisfy the extremal condition of the functional
\begin{align}
J &= \int d^6w_1 \tilde{g} + \nonumber \\
&\lambda_1\left(\int d^6 w_1 f^{(1)} - 1\right) + \lambda_2\left(\int d^6 w_1 \tilde{g} f^{(1)} -1\right) 
\end{align}
where the $\lambda_1$ and $\lambda_2$ terms are Lagrange multipliers enforcing normalization of $f^{(1)}$ and $\tilde{g}$. The extremal solutions are found to be of the form
\begin{equation}
\tilde{g} = \frac{C -\lambda_1 f_{+} }{1+ \lambda_2 f_+}
\end{equation}
where $C$ is constrained by the correlation present in the initial configuration. For instance, a large collection of electrons configured in a Fermi sphere model would have $C=1/2$. A separable Bose condensate gives $C=1$, with Lagrange multipliers also conforming to the uncorrelated  ``classical'' dynamics.

The extremal values of $\lambda_1$ and $\lambda_2$ may be found through the constraint equations, which reduce down to the single expression 
\begin{equation}
1 = \int d^6 w_1f^{(1)}(w_1,t) \frac{C-\lambda_1 f_+}{1+ \lambda_2 f_+} 
\end{equation}
where the maximal and minimal values distinguish respectively the boson and fermion solutions. The Lagrange multiplier solutions depend on global integrals of the DF, owing to the global reach of the permutation group. The multipliers are also integrals of motion, requiring only a single calculation at the initial conditions to set the XC strength. The dynamics of the condensate may then be written in a near-Boltzmann form 
\begin{align}
&0=\partial_t f^{(1)} + \frac{\vec{p}}{m} \cdot \vec{\nabla} f^{(1)} - \frac{N-1}{N} \vec{\nabla} \bar{\Phi} \cdot \vec{\nabla}_p f^{(1)} - \nonumber \\
& \frac{N-1}{N}\int d^6w_2 \vec{\nabla} \Phi_{12} \cdot \vec{\nabla}_{p} \Bigg(f^{(1)}(w_1,t) f^{(1)}(w_2,t) \times \nonumber \\
&\left(\frac{C-1 - \lambda_+ f_+}{1+\lambda_2 f_+}\right)\Bigg) \nonumber \\
\end{align}
where $\bar{\Phi}$ is the N-body averaged inter-particle potential, $\Phi_{12} = N \phi_{12}$, and $\lambda_+ = \lambda_1+\lambda_2$. The XC influence is encapsulated in the last term.

\section{\label{sec:Example}Demonstration}

In a simple demonstration of this method, consider the QCD axion, a dark matter candidate. Dark matter is expected to seed much of the large-scale structure of the visible universe through its gravitational collapse. The QCD axion is expected to be quite light, in the $\mu$eV-meV range, and it is expected to be extraordinarily degenerate at the presumed dark matter density and temperature \cite{Marsh2016,Guth2015,Sikivie2009}. Further, QCD axion dark matter is not expected to be significantly impacted by elements of the axion potential beyond the mass term on galactic scales and above, nor by interactions with the standard model such as the Primakov and chiral terms \cite{Marsh2016}.

A thin and cold spherical shell of matter is known to collapse under Newtonian self-gravitation, with expectations that the degenerate axion fluid be solely additionally impacted by XC on galactic scales. For simplicity, we have additionally assumed no tangential axion motion, that the radial velocity dispersion is sufficiently small so as to leave the shell width unchanged over the simulated collapse. These symmetries reduce the dimensionality of the problem from $7$ to $1$. Also, constraints of $C=1$, and $|\lambda_2 f_+| \ll 1$ are used. Finally, $\lambda_+$ is taken as a free tunable parameter in this demonstration. The governing equation of the Bose collapse may then be written to leading order in dispersion as
\begin{align}
\ddot{r} &= -\frac{GM}{r_{soft}^2} \nonumber \\
&+\lambda_+\left(\frac{3 G M N(r_{soft})}{4 r_{soft}^2} - \frac{G M N'(r_{soft})}{4 r_{soft}}\right),
\end{align}
where $M$ is the effective gravitating mass, $N(r_{soft})$ is the shaping coefficient of the shell distribution, defined by
\begin{equation}
N(r_{soft}) = \frac{1}{8 \pi^2 r_{soft}^2 \sigma_r \sigma_v},
\end{equation}
where $r_{soft}^2 = \left(r^2 + \sigma_r^2\right)$, $\sigma_r$ is the width of the shell, and $\sigma_v$ is the velocity dispersion of the shell. Note that the effective repulsion of the second term in the XC contribution will never dominate over the attraction of the first. Solutions of this system show that infall is amplified for non-trivial XC contributions, Fig. \ref{sph_shell_coll}. The new physics leads to a characteristically more violent infall due to the sharper form of the $\sim 1/r_{soft}^4$ central XC force. Parameters chosen for Fig. \ref{sph_shell_coll} are $r(t=0)=1$, $\sigma_r=10^{-3}$, and $\sigma_v=10^{-4}$ in dynamical units.

\section{\label{sec:Conclusions}Conclusions}

This communication derives the dynamics of non-relativistic condensed fluids with non-local inter-particle forces. It is found that the (anti-)symmetric condensed state accumulates sufficient strength to alter the system's dynamics on macroscopic scales. This model is related to the standard mean-field theory in the separable limit. A simple example of the model applied to QCD axion dark matter demonstrates notable deviation from classical gravitational collapse.

There are many other potential applications of this technique. The authors' own interests are in using the model for axion structure formation, but other fields such as condensed matter, nuclear astrophysics, cold atomic physics, and many others may find the above description helpful. An extension from the condensed limit to mixed systems of identical particles may be needed to describe axion dynamics. A letter detailing the expansion to fermions as well as bosons with non-local interactions is in preparation.

\section{\label{sec:Acknowledgements}Acknowledgements}

We would like to thank Jens Niemeyer, Katy Clough, David Marsh, Bodo Schwabe, Jan Velmatt, and Xiaolong Du for their productive discussions in the refinement of this paper. We also gratefully acknowledge the support of the U.S. Department of Energy office of High Energy Physics and the National Science Foundation. TQ was supported in part by the NSF grant AST-1514868. EL and LR were supported in part by the DOE grant DE-SC0011665.

\bibliographystyle{apsrev4-1}
\bibliography{Bibliography.bib}

\end{document}